\documentclass[aps,prd,showpacs,onecolumn,10pt]{revtex4}
\usepackage[colorlinks,linkcolor={green},citecolor={blue}]{hyperref}
\usepackage{eurosym}
\usepackage{graphicx}
\usepackage{amsmath}
\usepackage{amssymb}
\usepackage{amsfonts}
\usepackage{bm}
\usepackage{color}

\def\be{\begin{equation}}
\def\ee{\end{equation}}
\def\bea{\begin{eqnarray}}
\def\eea{\end{eqnarray}}

\newcommand{\f}[2]{\frac{#1}{#2}}

\begin{document}

\title{The Scalar Einstein-aether theory}
\author{Zahra Haghani$^1$}
\email{z.haghani@du.ac.ir}
\author{Tiberiu Harko$^2$}
\email{t.harko@ucl.ac.uk}
\author{Hamid Reza Sepangi$^3$}
\email{hr-sepangi@sbu.ac.ir}
\author{Shahab Shahidi$^1$}
\email{s.shahidi@du.ac.ir}
\affiliation{$^1$ School of Physics, Damghan University, Damghan,
41167-36716, Iran}
\affiliation{$^2$Department of Mathematics, University College London, Gower Street,
London WC1E 6BT, United Kingdom}
\affiliation{$^3$Department of Physics, Shahid Beheshti University, G. C., Evin, Tehran
19839, Iran}
\date{\today}

\begin{abstract}
We consider an Einstein-aether type Lorentz-violating theory of gravity in
which the aether vector field $V_{\mu }$ is represented as the gradient of
a scalar field $\phi $, $V_{\mu }=\nabla _{\mu }\phi $. A self
interacting potential for the scalar aether field is considered, as well as the
possibility of a coupling between the hydrodynamic matter flux and the aether
field, with the imposition of the timelike nature of the aether vector. The
gravitational field equations and the equation of motion of the scalar field are
derived by varying the action with respect to the metric and $\phi $. In the
absence of matter flux and scalar field coupling the effective energy-momentum
tensor of the scalar aether is conserved. The matter flux-aether coupling
generates an extra force acting on massive test particles and consequently the
motion becomes non-geodesic. The Newtonian limit of the theory is investigated
and the generalized Poisson equation for weak gravitational fields is obtained.
The cosmological implications of the theory is also considered and it is shown
that in the framework of the Scalar Einstein-aether theory both decelerating and
accelerating cosmological models can be constructed.
\end{abstract}

\maketitle

\section{Introduction}

In 1955 G. Szekeres \cite{Szek1}  proposed an extension of general
relativity in which the cosmic time is introduced as a new field variable.
The proposed formalism also allowed for a rigorous definition of the
concept of aether. In proposing such an extension of general relativity Szekeres
was motivated by the contradiction between the principle of equivalence,
 that all frames of reference are equivalent, and that there are no
physical processes which would distinguish one particular frame from the other,
and
the Weyl postulate, fundamental in cosmology, which requires the
existence of an absolute time. In order to give a rigorous definition of the
aether, Szekeres  introduced a scalar field variable $\phi $, called the cosmic
time and  postulated an interaction between fields associated with the metric
and the $\phi $
fields. The introduction of the cosmic time field $\phi $ allows for the
definition of the aether, which ``is a state of motion determined uniquely by
the gradient of $\tau $ at every point of the space time continuum" \cite%
{Szek1}. With the help of the gradient $Q_{\mu }$ of the cosmic time, $%
Q_{\mu}=\nabla _{\mu}\phi=\partial \phi/\partial x^{\mu}$, Szekeres
constructed the action of the gravitational field as
\begin{equation}  \label{1}
S=\int{\left\{R+\left[\frac{1}{2}\beta \left(\nabla ^{\nu}Q^{\sigma }\nabla
_{\nu}Q_{\sigma}\right)-\frac{\gamma }{\phi ^2}\right]+L_m\right\}\sqrt{-g}%
d^4x},
\end{equation}
where $R$ is the curvature scalar, $L_m$ the matter action, and $\beta $ and
$\gamma $ are constants. The term $-\gamma /\phi ^2$ in the
action describes the effect of the cosmological constant on the
gravitational dynamics. The gravitational field equations corresponding to
action (\ref{1}) were obtained in \cite{Szek1} and their physical
implications (cosmological models, spherically symmetric vacuum solution,
planetary orbits, gravitational energy and gravitational waves) were
investigated in detail. An extension of the aether model was considered in
\cite{Szek2}, where a more general Lagrangian of the form $L=R+\gamma
_1S_1+\gamma _2S_2$ was constructed with $\gamma _1$,  $\gamma _2$ being
constants and $%
S_1$,  $S_2$ are scalar densities formed from the cosmic time field $\phi
$ and the metric tensor components, representing the energy density of the
cosmic time and the interaction of the $\phi $ and $g_{\mu \nu}$ fields,
respectively. The authors adopt $S_1\sim \phi ^{-2}$,  and consider two choices
for $S_2$, $%
S_2=\left(C_{\mu}^{\mu}\right)^2$ and $S_2=\left(C^{\nu}_{\mu}C^{\mu
}_{\nu}\right)$, respectively, where $C_{\mu \nu}=\nabla _{\nu}Q_{\mu}$. The
physical implications of this model were also analyzed in detail. The
gravitational motion in the presence of aether drift was considered in \cite%
{Szek3}. It was shown that the path of a free particle is geodesic provided
that its absolute velocity is small compared to the velocity of light.
Hence the aether drift does not have a direct effect on the motion of
particles in slow motion. However, it influences and modifies the external
gravitational field of a massive body and the associated geodesics. In the
case of the field generated by a spherically symmetric object, due to the
action of the aether, some asymmetry is present which makes it possible to
detect and measure the aether drift vector $\partial
\phi/\partial x^{\mu }$ experimentally. The effect is small, of the same
order of magnitude as that of the general relativistic corrections to the
Newtonian
theory, with the velocity of aether drift in the Solar System being of the
order of
100 km/s.

Recently T. Jacobson has proposed a Lorentz-violating theory of
gravity with an
``aether" vector field $V_{\mu }$, determining a preferred
rest frame at each space-time point; the so-called
Einstein-aether (EA) gravity theory \cite{Jac1}. More precisely, $V_{\mu }$
breaks local
boost invariance, while rotational symmetry in a preferred frame is
preserved \cite{Jac2}.
The most general action for the pure EA theory is given by
\cite{aet}
\begin{align}\label{act}
S_{ae}=\frac{1}{2\kappa^2}\int d^4x\sqrt{-g}\Big[R+K_{\lambda \sigma }^{\mu
\nu }\nabla _{\mu } V^{\lambda }\nabla _{\nu } V^{\sigma }+\lambda \left(V_{\mu
} V^{\mu }+1\right)\Big]+S_m,
\end{align}
where the Lagrange multiplier $\lambda $ enforces the  constraint on
the  vector field $V_{\mu }$  being timelike. The tensor $K^{\mu \nu}
_{\lambda
\sigma}$ is given by
\begin{equation}\label{coef}
K^{\mu \nu}_{\lambda \sigma } = c_0g^{\mu \nu }g_{\lambda \sigma } +
c_1\delta ^{\mu } _{\sigma } \delta ^{\nu } _{\lambda }+ c_2\delta ^{\mu
}_{\lambda } \delta ^{\nu }_{\sigma } + c_3V^{\mu }V^{\nu } g_{\lambda
\sigma },
\end{equation}
where $c_i$, $i=0,1,2,3$ are the dimensionless free parameters of the
EA theory. The action given by Eq.~(\ref{act}) extends the standard Einstein-Hilbert
action  for the metric with the addition of a kinetic term for the aether,
containing four dimensionless coefficients $c_{i}$, $i=0,3$ which couple the
aether to the metric through the covariant derivatives and a non-dynamical
Lagrange multiplier $\lambda $.

One of the other representations of a Lorentz violating extension of general
relativity was proposed by P. Horava \cite{hora}. The Horava-Lifshitz (HL)
theory was written as an attempt to build a UV completion of general relativity
by adding higher order spatial derivatives to the theory without adding higher
order time derivatives. This results in modification of the
graviton propagator in such a way that the theory becomes power counting
renormalizable. Horava assumed a preferred space-like foliation of space-time
which can be described by a scalar field and the lapse function $N$ which depends
only on time in the projectable version of the theory. This restriction
implies that all spatial derivatives of $N$ vanish. Also many different terms
become identical up to a total derivative which makes the calculations
tractable. However, it was shown that this scalar mode for gravity causes
problems such as  instability and strong coupling at low energies \cite{inst}. This problem
can be avoided if one adds all the possible terms which respect the symmetry of
the theory to the action \cite{BPS}. One can
then obtain a theory where its strong coupling scale is pushed to sufficiently
high energies.     The theory can be considered as an effective field theory which we
denoted by BPS theory.

The HL and the EA theories are both modifications of
gravity which break the Lorentz symmetry. This suggests that these theories
may be related to each other. In fact, in the limit where higher than second
order derivative terms of the HL theory can be ignored (which
corresponds to the IR limit of the theory), one obtains the EA
theory with an additional constraint that the aether vector should be
hypersurface orthogonal \cite{BPS,HL}. Moreover, because all spherically
symmetric solutions are hypersurface orthogonal, one can expect that all these
solutions of EA theory should also be a solution to the IR limit
of HL theory \cite{black}. The above arguments suggest that the
EA theory can be seen as a covariant version of the IR limit of
HL gravity. In other worlds, the EA theory, and as
a consequence the present theory, can be motivated by the fact that it comes
from the IR limit of a UV complete theory of gravity. In order to go further,
we note that in the BPS theory one substitutes a fixed spatial
foliation by a dynamical one. So, the theory can be written in a covariant way,
by using an additional scalar field $T$ \cite{ted}. If one inserts this scalar field
to the theory as a dynamical variable, then the general covariance can be
preserved.
This scalar field can be related to the aether vector of EA
theory by $u_\mu=W\nabla_\mu T$ where $W$ is a normalization factor, which is
defined as
    $$W=(g^{\mu\nu}\nabla_\mu T\nabla_\nu T)^{1/2}.$$
If this vector field is also hypersurface orthogonal, then the above
EA-restricted theory is identical to BPS theory.
We should mention that our work puts another restriction on the above
EA-restricted theory, which is the condition that the $W$ field
should also be proportional  to $T$ itself and not to its derivative as in the above theory
\cite{tedi}.

An interesting alternative dark matter model was introduced by Milgrom
\cite{MOND} in which Newton's second law is modified for very small
accelerations, commonly known as MOND. A relativistic version of MOND was
proposed by Bekenstein \cite{Bek} where gravity is mediated by three fields, a
tensor field with an associated metric compatible connection, a timelike one
form field, and a scalar field respectively. However, in \cite{Zlos} it was
shown that Bekenstein's theory can be reformulated  as a Vector-Tensor theory
akin to EA theory with non-canonical kinetic terms. The total
TeVeS action can be entirely written in the matter frame, and is given by
\bea
S_{TeVeS}=\frac{1}{16\pi G}\int\Bigg[R+\tilde{K}^{abmn}\nabla _aA_m\nabla
_bA_n+\frac{V(\mu)}{A^2}\Bigg]\sqrt{-g}d^4x+S_m\left[g^{ab}\right],
\eea
where $\mu $ is a non-dynamical field.

The physical and cosmological implications of the EA type theories
have been intensively investigated. Time independent spherically
symmetric
solutions of this theory were studied in \cite{aet1} and a three-parameter
family of solutions was found. Asymptotic flatness restricts the solutions to a
two parameter class and requiring the aether to be aligned with the timelike
Killing field further restricts them to one parameter, the total mass. The
static aether solutions are given analytically up to the solution of a
transcendental equation. Black Holes in EA theory were studied in
\cite{aet2}. To be causally isolated, a black hole interior must trap matter
fields as well as all aether and metric modes. The theory possesses spin-0,
spin-1, and spin-2 modes whose speeds depend on the four coupling coefficients.
The gravitational spectrum of black holes in the EA theory was
considered in \cite{aet3}, while numerical simulations of the gravitational
collapse, neutron star structure, strong field effects and generic properties
of black holes were analyzed in \cite{aet4}. Post-Newtonian approximations,
solar system and galactic and extra-galactic tests of the theory were obtained
and discussed in \cite{aet5}.  By coupling a scalar field to the timelike vector
in \cite{aet5a} it was shown via a tunneling approach that the universal
horizon radiates as a black body at a fixed temperature even if the scalar
field equations also violate local Lorentz invariance.
A comprehensive study of the cosmological effects of the
EA theory was performed in \cite{aet6a} and  observational data
were used to constrain it. In conjunction with the previously determined
consistency and experimental constraints, it was found that an EA
universe can fit observational data over a wide range of its parameter space
but requires a specific re-scaling of the other cosmological densities. Another
intersting application of aether theory in cosmology has been discussed in
\cite{24} where the authors proposed a new class of theories where energy
always flows along timelike geodesics, mimiking dark energy.

The primordial perturbations generated during a stage of single-field inflation
were analyzed in \cite{aet7}.  Quantum fluctuations of the inflaton and aether
fields would seed long wavelength adiabatic and isocurvature scalar
perturbations, as well as transverse vector perturbations. Scalar and vector
perturbations may leave significant imprints on the cosmic microwave
background. The primordial spectra and their contributions to temperature
anisotropies were obtained and some of the phenomenological constraints that
follow from observations were formulated.  The linear perturbation equations
were constructed in a covariant formalism and  the CMB B-mode polarization,
using the CAMB code, was modified so as to incorporate the effects of the aether
vector field  \cite{aet8}. Several families of accelerating universe solutions
to an EA gravity theory were derived in \cite{aet9}. These
solutions provide possible descriptions of inflationary behavior in the early
universe and late-time cosmological acceleration. By taking a special form of
the Lagrangian density of the  aether field  it was shown in \cite{aet10} that
the EA theory may represent an alternative to the standard dark
energy model.  A dynamical systems analysis to investigate the future behavior
of EA cosmological models with a scalar field coupling to the
expansion of the aether and a non-interacting perfect fluid was performed in
\cite{aet11}. The stability of the equilibrium solutions were analyzed and the
results were  compared to the standard inflationary cosmological solutions
and previously studied cosmological EA models.  A class of
spatially anisotropic cosmological models in EA theory with a
scalar field in which the self-interacting potential depends on the timelike
aether vector field through the expansion and shear scalars was investigated in
\cite{aet12}. The cosmological evolution of EA models with a
power-law like potential,  using the method of dynamical systems, was studied in
\cite{aet13}. In the absence of matter, there are two attractors which
correspond to an inflationary universe in the early epoch, or a de Sitter
universe at late times. In the case where matter is present, if there is no
interaction between dark energy and matter, there are only two de Sitter
attractors and no scaling attractor exists. The consequences of Lorentz
violation during slow-roll inflation were analyzed in \cite{aet14}.  If the
scale of Lorentz violation is sufficiently small compared to the Planck mass and
the strength of the scalar-aether coupling is suitably large, then the spin-0
and spin-1 perturbations grow exponentially and spoil the inflationary
background. The effects of such a coupling on the Cosmic Microwave Background
(CMB) are too small to be visible to current or near-future CMB experiments.

It is the goal of the present paper to consider a scalar formulation of the
EA theory where the aether four-vector $V_{\mu }$ can be
represented as the
gradient of a scalar function, so that $V_{\mu }=\nabla _{\mu }\phi$. Therefore
we consider a gravitational theory where the standard Einstein-Hilbert action
is modified by considering  aether kinetic terms which are coupled to the
metric via the coupling coefficients given by Eq.~(\ref{coef}). Moreover, we
impose the timelike constraint on the gradient of the aether scalar field via a
Lagrange multiplier. A self-interacting potential for the scalar field is then
included in the action. In addition, we  consider the possibility of a
coupling/interaction between the matter flux and the gradient of the scalar
field. In this way the matter in motion would feel the effects of the aether
which can directly influence the dynamics of massive test particles. The
gravitational field equations corresponding to the scalar EA
action are obtained by varying the metric as well as the scalar field. In the
absence of matter current-aether field coupling the covariant divergence of the
effective energy-momentum tensor corresponding to the aether field is zero. The
matter flux-aether field coupling induces  non-conservation of the
energy-momentum tensor. The corresponding particle motion is non-geodesic and
the equation of motion of a massive test particle is obtained for this case.
The Newtonian limit of the theory and the generalized Poisson equation is then
derived. We briefly consider the cosmological implications of the theory and
show that, depending on  the values of  coupling constants $c_i$, $i=0,1,2,3$, a
large number of cosmological solutions, both accelerating and decelerating, can
be obtained.

The present paper is organized as follows. The gravitational field equations of
the scalar EA theory are presented in Section~\ref{sect1}. The
equation of motion of massive test particles, the Newtonian limit of the theory
and the generalized Poisson equation are obtained in Section~\ref{sect2}. The
cosmological implications of the theory are considered in Section~\ref{sect3}.
We discuss and conclude our results in Section~\ref{sect4}. The computation of
the divergence of the energy-momentum tensor of the theory is presented in
detail in the Appendix.

\section{Gravitational field equations of the scalar EA
theory}\label{sect1}

The vector EA theory is defined by the action given by
Eq.~(\ref{act}).
In the following we consider that the aether vector field $V_{\mu}$ can be
represented as the gradient of a scalar function, that is
\begin{equation}
V_{\mu }=\nabla _{\mu }\phi.
\end{equation}
After such substitution, it turns out that the terms with coefficients $c_0$
and $c_1$ becomes equal. We then propose the scalar EA (SEA)
gravitational theory
action as
\begin{align}  \label{ac1}
S_{SEA}=\frac{1}{2\kappa^2}&\int d^4x\sqrt{-g}\Big[ R+c_1\nabla_\mu\nabla_\nu%
\phi\nabla^\mu\nabla^\nu\phi+c_2(\Box\phi)^2 +c_3\nabla^\mu\phi
\nabla^\nu\phi\nabla_\mu\nabla_\sigma\phi\nabla_\nu\nabla^\sigma\phi+c_4\rho
u^\sigma\nabla_\sigma\phi  \notag \\
&-V(\phi)+\lambda(\nabla_\mu\phi\nabla^\mu\phi +\epsilon)\Big]+S_m,
\end{align}
where $\epsilon =\pm1$, and $c_4$ is a constant. For $\epsilon =1$ we assume
that, analogous  to the vector EA case, the scalar function $%
\phi $ is normalized via $\nabla _{\mu }\phi \nabla ^{\mu }\phi =-1$. The
term $\rho u^{\sigma }\nabla _{\sigma }\phi$ represents a possible
interaction between the matter hydrodynamic flux $j^{\sigma }=\rho
u^{\sigma} $ and the aether vector. We have also added to the action a
self-interacting scalar field potential $V(\phi)$.

One can also write the
above action in a compact way as
\begin{align}
S_{SEA}=\frac{1}{2\kappa^2}\int d^4x\sqrt{-g}\Big[ R+(\nabla_\mu\nabla_\nu%
\phi)K^{\mu\nu,\rho\sigma}(\nabla_\rho\nabla_\sigma\phi) +c_4\rho
u^\sigma\nabla_\sigma\phi-V(\phi)+\lambda(\nabla_\mu\phi\nabla^\mu\phi
+\epsilon)\Big]+S_m,
\end{align}
where we have defined
\begin{align}
K^{\mu\nu,\rho\sigma}=c_1g^{\mu\rho}g^{\nu\sigma}+c_2g^{\mu\nu}g^{\rho%
\sigma}+c_3g^{\nu\sigma}\nabla^\mu\phi\nabla^\rho\phi.
\end{align}

Varying action \eqref{ac1} with respect to the Lagrange multiplier $%
\lambda$ we immediately find that
\begin{align}  \label{eqlam}
\nabla_\mu\phi\nabla^\mu\phi=-\epsilon,
\end{align}
which introduces a preferred direction for the space-time. Varying with
respect to the metric leads to the equation of motion
\begin{widetext}
\begin{align}\label{eq-1}
G_{\mu\nu}+K_{\mu\nu}+\f{1}{2}V(\phi)g_{\mu\nu}
+\lambda\nabla_\mu\phi\nabla_\nu\phi=\kappa^2 T_{\mu\nu}
\end{align}
where we have used equation \eqref{eqlam} to simplify the above equation and we
have defined
\begin{align}
K_{\mu\nu}
&=c_1\bigg(\nabla^\lambda(\nabla_\lambda\phi\nabla_\mu\nabla_\nu\phi)
-2\Box\nabla _ { (\mu }
\phi\nabla_{\nu)}
\phi-\f{1}{2}g_{\mu\nu}
\nabla_\alpha\nabla_\beta\phi\nabla^\alpha\nabla^\beta\phi\bigg)\nonumber\\
&+c_2\bigg(g_{\mu\nu}\nabla^\lambda\phi\nabla_\lambda\Box\phi+\f{1}{2}g_{\mu\nu}
(\Box\phi)^2-2\nabla_{(\mu}\phi\nabla_{\nu)}\Box\phi\bigg)
+\f{1}{2}c_4T_{\mu\nu}u^{\alpha}\nabla_\alpha\phi.
\end{align}

The equation of motion for the aether scalar becomes
\begin{align}\label{eq-2}
c_1\nabla_\nu\Box\nabla^\nu\phi+c_2\Box^2\phi-\f{1}{2}c_4\nabla_\mu\left(\rho
u^\mu\right)-\f{1}{2}\f{\textmd{d}V}{\textmd{d}\phi}
-\nabla_\mu\left(\lambda\nabla^\mu\phi\right)=0.
\end{align}
\end{widetext}
One should note that the $c_3$ term does not contribute to the equations of
motion. This is because we have used the constraint equation \eqref{eqlam} and
its derivative
\begin{align}\label{1}
\nabla_\mu\nabla^\alpha\nabla^\mu\phi=0.
\end{align}
We also note that in the special case of $V(\phi)=0$, the scalar equation
becomes a total derivative $\nabla_\nu J^\nu=0$, with
\begin{align}\label{2}
J^\nu=c_1\Box\nabla^\nu\phi+c_2\nabla^\nu\Box\phi-\f12 c_4\rho
u^\nu-\lambda\nabla^\nu\phi,
\end{align}
which gives $\sqrt{-g}J^\nu=const.$ Assuming that the constant being zero, and
Multiplying the whole equation by $\nabla_\nu\phi$, one can obtain the Lagrange
multiplier $\lambda$ as
\begin{align}\label{3}
\epsilon\lambda=\f12 c_4\rho
u^\nu\nabla_\nu\phi-c_1\nabla^\nu\phi\Box\nabla_\nu\phi-c_2\nabla_\nu\phi\nabla^
\nu\Box\phi.
\end{align}
In the following we assume that the matter content
of the Universe is represented by a perfect fluid with energy density $\rho $
and thermodynamic pressure $p$, with energy-momentum tensor given by
\begin{equation}  \label{ene}
T_{\mu }^{\nu}=\left(\rho +p\right)u^{\mu}u_{\nu}+p\delta _{\mu}^{\nu},
\end{equation}
where $u^{\mu}$ is the velocity four-vector of the matter.
Using the equations of motion, one can prove that the terms proportional to
$c_1$%
, $c_2$ and $c_3$ do not contribute to the covariant derivative of the ordinary
matter
energy-momentum tensor. One can then obtain (see the Appendix for details) the
covariant divergence of the matter energy-momentum tensor as
\begin{align}  \label{eq-3}
\nabla^\mu T_{\mu\nu}=\frac{c_4\big[T_{\mu\nu}\nabla^\mu(u^\alpha\nabla_%
\alpha\phi) -\nabla_\alpha(\rho u^\alpha)\nabla_\nu\phi\big]}{%
2\kappa^2-c_4u^\beta\nabla_\beta\phi}
\end{align}
If $c_4=0$, that is, we neglect the possible coupling between the matter flux
$j^{\sigma }$ and the aether vector, the matter energy-momentum is conserved,
$\nabla^\mu T_{\mu\nu}=0$.

\section{Equation of motion of a massive test particle and the Newtonian
limit}\label{sect2}

In this Section we obtain the equation of motion for a massive test particle
moving in a SEA universe. The Newtonian limit of the
theory is considered and the generalized Poisson equation for the gravitational weak field
is derived.

\subsection{The equation of motion of massive test particles}
 Taking the divergence of Eq.~\eqref{ene}
and defining the projection operator $h_{\mu\nu}=g_{\mu\nu}+u_\mu u_\nu$ one
obtains
\begin{align}
\nabla_\mu T^{\mu\nu}=h^{\mu\nu}\nabla_\mu p + u^\nu u_\mu\nabla^\mu \rho
+(\rho+p)\big(u^\nu\nabla_\mu u^\mu+u^\mu\nabla_\mu u^\nu\big).
\end{align}
Multiplying the equation above by $h^\lambda_\nu$ we have
\begin{align}
h_\nu^\lambda\nabla_\mu T^{\mu\nu}=(\rho+p)u^\mu\nabla_\mu
u^\lambda+h^{\nu\lambda}\nabla_\nu p,
\end{align}
where we have used the relation $u_\mu\nabla_\nu u^\mu=0$. Noting that
\begin{align}  \label{em-3}
u^\mu\nabla_\mu u^\lambda=\frac{d^2x^\lambda}{ds^2}+\Gamma^\lambda_{~\mu%
\nu}u^\mu u^\nu,
\end{align}
and using Eq.~\eqref{eq-3} we obtain the equation of motion for a
massive test particle as
\begin{align}  \label{em-4}
\frac{d^2x^\lambda}{ds^2}+\Gamma^\lambda_{~\mu\nu}u^\mu u^\nu=f^\lambda,
\end{align}
where
\begin{align}  \label{em-5}
f^\lambda=\frac{h^{\lambda\nu}}{\rho+p}\left(\frac{c_4\big[
\nabla_\nu(pu^\alpha\nabla_\alpha\phi)
-\nabla_\nu\phi\nabla_\alpha(\rho u^\alpha)\big]-2\kappa^2\nabla_\nu p}{%
2\kappa^2-c_4 u^\beta\nabla_\beta\phi}\right).
\end{align}
$f^\lambda$ is the extra force which leads to non-geodesic motion for a
massive test particle in the SEA universe. We note that if $c_4$ vanishes, the
above equation reduces to the standard geodesic equation for a perfect
fluid. We also note that the extra force is perpendicular to the particle
four-velocity, $f^{\nu}u_\nu=0$.

\subsection{The Newtonian Limit of the SEA theory}

In order to obtain the Newtonian limit of the theory we first show that the
equation of motion Eq.~\eqref{em-4} can also be obtained by a variational
principle. We assume that the extra force given by Eq.~\eqref{em-5} can be
written
formally as
\begin{align}  \label{nl-1}
f^{\lambda} = (g^{\nu \lambda}+u^{\nu}u^{\lambda})\nabla_{\nu} \ln \sqrt{Q},
\end{align}
where $Q$ is a dimensionless quantity. We note that when $Q$ tends to unity
we recover the standard geodesic equation of general relativity. Now, in order
to obtain the form
of $Q$ in the Newtonian limit of SEA theory we assume that the density of
the physical system is small and one may ignore the pressure $p\ll\rho$. In
this case, using Eqs.~\eqref{em-5} and \eqref{nl-1}, one has
\begin{align}\label{n1-2}
\nabla_\nu\ln\sqrt{Q}=\frac{1}{\rho}\frac{c_4\nabla_\nu\phi\nabla_\alpha(%
\rho u^\alpha)}{c_4u^\beta\nabla_\beta\phi-2\kappa^2}.
\end{align}
In the Newtonian limit the function $\phi$ depends only on $r$ and the
velocity of the particle satisfies $u^{\mu}=\delta^\mu_0/\sqrt{g_{00}}$. One
can then easily show that $u^\beta\nabla_\beta\phi\approx0$. We can also
assume that $\phi=\phi(\rho)$ and
\begin{align}  \label{nl-3}
\nabla_\alpha(\rho u^\alpha)\equiv Z(\rho).
\end{align}
Expanding Eq.~\eqref{n1-2} about the background density $\rho_0$,
one has
\begin{align}  \label{nl-4}
\nabla_\nu\ln\sqrt{Q}=-\frac{c_4}{2\kappa^2}\phi^\prime(\rho_0)\bigg[\frac{%
Z(\rho_0)}{\rho}+Z^\prime(\rho_0)\bigg]\nabla_\nu\rho.
\end{align}
The first term of the above equation is proportional to $\nabla_\nu\ln\rho$,
which can be further simplified to
\begin{align}  \label{nl-5}
\nabla_\nu\ln\rho=\nabla_\nu\ln(\rho_0+\delta\rho)\approx \frac{1}{\rho_0}%
\nabla_\nu\delta\rho,
\end{align}
where $\delta\rho=\rho-\rho_0$. One may then obtain the following expression
for $\sqrt{Q}$
\begin{align}  \label{nl-6}
\sqrt{Q}\approx 1-\frac{\alpha}{\rho_0}\delta\rho,
\end{align}
where we have defined
\begin{align}  \label{nl-7}
\alpha=\frac{c_4}{2\kappa^2}\phi^\prime(\rho_0)\big[Z(\rho_0)+\rho_0
Z^\prime(\rho_0)\big].
\end{align}
We note that one may obtain the equation of motion Eq.~\eqref{em-4} by starting
with the formal definition of the extra-force given in
Eq.~\eqref{nl-1}, and varying the modified particle motion action \cite{22fRT}
\begin{align}  \label{n1-8}
S_p=\int\sqrt{Q}\sqrt{g_{\mu\nu}u^\mu u^\nu}ds.
\end{align}
One can see that in the case $\sqrt{Q}\rightarrow1$, the standard geodesic
equation is obtained. In the Newtonian limit the standard line element
for a dust  particle can be written as
\begin{align}  \label{nl-9}
\sqrt{g_{\mu\nu}u^\mu u^\nu}ds\approx\bigg(1+\psi-\frac{\vec{v}^2}{2}\bigg)%
dt,
\end{align}
where $\psi$ is the Newtonian potential and $\vec{v}$ is the 3D velocity of
the particle. Using Eqs.~\eqref{nl-6} and \eqref{nl-9} one can write the action
\eqref{n1-8} as
\begin{align}  \label{nl-10}
S_p=\int\bigg(1+\psi-\frac{\vec{v}^2}{2}-\frac{\alpha}{\rho_0}\delta\rho\bigg)dt.
\end{align}
Varying the above action gives us
\begin{align}  \label{nl-11}
\vec{a}=-\vec{\nabla}\psi+\vec{a}_E,
\end{align}
where the first term is the Newtonian acceleration $\vec{a}_N$ and the
second term is the extra acceleration having the form
\begin{align}  \label{nl-12}
\vec{a}_E(\rho)=\frac{\alpha}{\rho_0}\vec{\nabla}\rho.
\end{align}
The extra acceleration depends on the gradient of the matter density. The above
theory shows that in the region of space-time where  the matter density is
(approximately) constant, the extra acceleration
becomes zero or  is negligibly small.

\subsection{The generalized Poisson equation}

Taking the trace of the gravitational field equation, Eq.~\eqref{eq-1}, and
assuming that $V(\phi)=0$, one has
\begin{align}\label{4}
R=(c_1+2c_2)(\Box\phi)^2+(c_1+3c_2)\nabla^\nu\phi\nabla_\nu\Box\phi
+c_1\nabla^\nu\phi\Box\nabla_\nu\phi -c _ 4 \rho
u^\alpha\nabla_\alpha\phi+\kappa^2\rho.
\end{align}
One may easily obtain the solution
\begin{align}\label{5}
g_{\mu\nu}=\eta_{\mu\nu},\qquad \phi_0=2t+x+y+z,\qquad \lambda=0,
\end{align}
as a background solution of the theory. One should note that, because of the
constraint equation \eqref{eqlam}, the aether scalar should depend on time. So
one cannot obtain a fully static solution as a background solution.

Assuming that the metric takes the form
\begin{align}\label{6}
ds^2=-(1-2A(\vec{x}))dt^2+(1-2\psi(\vec{x}))(dx^2+dy^2+dz^2),
\end{align}
and
\begin{align}\label{7}
\phi=\phi_0+\varphi(\vec{x}),
\end{align}
one can see that the constraint equation \eqref{eqlam} can be satisfied, to 
first order if
\begin{align}\label{8}
\psi=\f43A-\f13\sum_{i=1}^3\partial_i\varphi.
\end{align}
One can then obtain, up to first order
\begin{align}  \label{new-1}
R&=\f{22}{3}\nabla^2A-\f43\sum\nabla^2\partial_i\varphi.
\end{align}
The generalized Poisson equation for the SEA theory now reads
\begin{align}  \label{new-2}
\nabla^2A=\f{2}{11}\sum\nabla^2\partial_i\varphi+\f{3}{22}(c_1+3c_2)\nabla^\nu
\phi_0\nabla_\nu\Box\varphi+\f{3}{22}c_1\nabla^\nu\Box\nabla_\nu\varphi+\f{3}{22
}(\kappa^2-2c_4)\rho.
\end{align}

\section{Cosmological Solutions}\label{sect3}

In this Section we will study the cosmological implications of the
SEA theory. We will restrict our study to homogeneous and
isotropic cosmological models, with the line element given by the flat
Friedmann-Robertson-Walker metric
\begin{align}  \label{cos-1}
ds^2=-dt^2+a(t)^2d\vec{x}^2.
\end{align}
We also assume that the scalar aether field is homogeneous and therefore has
the form $\phi=\phi(t)$. We also assume that the matter content
of the universe has a perfect fluid form
\begin{align}\label{cos-1.5}
T^\mu_{\nu}=\textmd{diag}\big[-\rho(t),p(t),p(t),p(t)\big],
\end{align}
and the velocity of the particle is $u^\alpha=[1,0,0,0]$.
In this case the constraint equation Eq.~\eqref{eqlam} can be solved for $\phi$
to
give
\begin{align}  \label{cos-2}
\phi=t+\alpha_1,
\end{align}
where $\alpha_1$ is an integration constant and we have assumed that $%
\epsilon=1$ so that the aether vector becomes a time-like vector field. With
these assumptions, equation \eqref{cos-2} becomes obvious since the FRW metric
already has a time-like preferred direction $\partial/\partial t$ and the
aether vector should be
identical to that direction up to a shift.
The metric and scalar field equations can then be obtained from %
\eqref{eq-1} and \eqref{eq-2} as
\begin{align}  \label{cos-3}
&3\big[3(c_1+c_2)-2\big]H^2-6c_2\dot{H}+(2\kappa^2-c_4)\rho-2\lambda=0, \\
&(c_1+3c_2-2)(3H^2+2\dot{H})-(2\kappa^2-c_4)p=0,
\end{align}
and
\begin{align}  \label{cos-4}
c_2\ddot{H}-(2c_1-3c_2)H\dot{H}-3c_1H^3+\frac{1}{3}\dot\lambda-\frac{1}{6}%
c_4\dot\rho+\frac{1}{2}(2\lambda-c_4\rho)H=0,
\end{align}
where we have assumed that $V(\phi)=0$. The conservation equation \eqref{eq-3}
reduces to
\begin{align}  \label{cos-5}
2(c_4-\kappa^2)(\dot\rho+3H\rho)+3(c_4-2\kappa^2)Hp=0.
\end{align}

\subsection{Vacuum solutions}

In the case of zero energy-momentum tensor $T_{\mu\nu}=0$ one can easily
obtain the dust-like solution $a=a_0t^{2/3}$, $a_0={\rm constant}$,  with the
Lagrange multiplier
\begin{align}  \label{cos-6}
\lambda=\frac{2(3c_1+6c_2-2)}{3t^2}.
\end{align}
Now, let us consider the case
\begin{align}  \label{cos-7}
c_2=\frac{2-c_1}{3},
\end{align}
which simplifies the equations. In this case one obtains a
self-accelerating solution $a=a_0\exp\left(H_0t\right)$, $H_0={\rm constant}$
with
\begin{align}  \label{cos-8}
\lambda=3c_1H_0^2.
\end{align}
There is also a power-law solution $a=a_0t^n$ with the Lagrange multiplier
of the form
\begin{align}  \label{cos-9}
\lambda=\frac{n\big[2+c_1(3n-1)\big]}{t^2}.
\end{align}

\subsection{The case $c_4=0$}

In this case, the energy-momentum tensor becomes conserved due to
equation \eqref{eq-3}. One may then obtain the matter dominated solution $%
a=a_0t^{2/3}$, $a_0={\rm constant}$
\begin{align}  \label{cos-10}
p=0,\quad \rho=\frac{\rho_0}{t^2},\quad \lambda=\frac{\lambda_0}{t^2},
\end{align}
with
\begin{align}  \label{cos-11}
\lambda_0=2c_1+4c_2-\frac{4}{3}+\rho_0\kappa^2.
\end{align}
One can also obtain a radiation solution $a=a_0t^{1/2}$ with
\begin{align}  \label{cos-12}
p=\frac{\rho_0}{3t^2},\quad
\rho=\f{\rho_0}{t^2}+\f{\rho_1}{t^{3/2}},\quad\lambda=\frac{3(2c_2+c_1)}{4t^2}
+\f{\kappa^2\rho_1}{t^{3/2}},
\end{align}
where
\begin{align}  \label{cos-13}
\rho_0=\frac{3(2-c_1-3c_2)}{8\kappa^2}.
\end{align}%
and $\rho_1$ is an integration constant.
\newline

\subsection{The case $c_4\neq0$}

In this case one has the matter-dominated solution $a=a_0t^{2/3}$, $a_0={\rm
constant}$.
Putting $p=0$ in \eqref{eq-3} one may prove that the energy-density should
behave as
\begin{align}  \label{cos-14}
\rho=\frac{\rho_0}{t^2},
\end{align}
as in the case where one has energy-momentum conservation. In this case we
obtain
\begin{align}  \label{cos-15}
\lambda=\frac{(6\kappa^2-3c_4)\rho_0-8+24c_2+12c_1}{6t^2}.
\end{align}
We have also a self-accelerating solution $a=a_0e^{H_0t}$, $H_0={\rm
constant}$, with
\begin{align}  \label{cos-16}
p&=\frac{3H_0^2(2-c_1-3c_2)}{c_4-2\kappa^2},  \notag \\
\rho&=\frac{3H_0^2(c_1+3c_2-2)}{2(c_4-\kappa^2)}+\rho_0e^{-3H_0t},  \notag \\
\lambda&=\frac{1}{2}(2\kappa^2-c_4)\rho+\frac{3}{2}H_0^2\big[3(c_1+c_2)-2%
\big].
\end{align}
The above equation shows that the energy-density has a cosmological constant
part. In order to get rid of the cosmological constant term one should again
impose
the condition \eqref{cos-7} which would make  the solution  a
self-accelerating one with an energy-momentum tensor behaving
as dust.

\section{Discussion and final remarks}\label{sect4}

In this paper we have considered a scalar version of the vector
EA type theories where the aether vector field is represented by
the gradient of a scalar function. In this way the basic physical
characteristics of the aether can be described in terms of a single scalar
function $\phi $, whose coupling to the metric is accomplished via its
gradient. A self-interacting potential of the scalar aether field can also be
added to the theory as well as a possible coupling between the hydrodynamic flow
of the matter, described by the flux $j^{\sigma }$, and the aether scalar. In
the presence of such a coupling the energy-momentum tensor of the matter is not
conserved and an extra force is generated. If no such coupling  exists, the
matter energy-momentum tensor is conserved since the covariant divergence of
the effective energy-momentum tensor of the scalar aether is identically zero.
In the presence of the matter flow-aether coupling, in the weak field limit,
the total acceleration of a massive test particle is
$\vec{a}=\vec{a}_N+\vec{a}_E$. As shown in \cite{Bert}, the Newtonian
acceleration can be expressed as $\vec{a}_N\approx
\left(|\vec{a}|/2|\vec{a}_E|\right)\vec{a}$, or, equivalently,
$a=\sqrt{2a_Ea_N}$, a relation which is very similar to the acceleration
equation introduced in the MOND approach to dark matter \cite{MOND}. Since
$a_N=GM/r^2$, where $M$ is the mass of the cental body, it follows that
$a\approx \sqrt{a_EGM}/r=v_{tg}^2/r$,  where $v_{tg}$ is the rotational velocity
of a massive test particle under the influence of a central force. Therefore, it
follows that $v_{tg}^2\rightarrow v_{\infty}^2=\sqrt{a_EGM}={\rm constant}$,
pointing to the presence of an extra force due to the coupling between
hydrodynamic motion and aether which may explain the constancy of the galactic
rotation curves, usually attributed to the presence of dark matter.

We have also investigated the cosmological implications of the theory by
studying the cosmological evolution of a flat, homogeneous and isotropic
Universe. Depending on the values of the coupling constants $c_i$, $i=0,1,2,3$,
several classes of cosmological models can be constructed. For simplicity, in
our analysis we have neglected the possible physical effects of the scalar
field self-interacting potential $V$.  In particular, a de Sitter type
cosmological expansion is possible in the presence of a hydrodynamic flow-scalar
field interaction. In this scenario, in the long time limit, the matter
energy-density tends to of a constant value. Power law solutions can be obtained
for dust and radiation filled Universes as well.

In conclusion, we have proposed a scalar version of the EA type
theories.
The scalar version of the EA theory was also considered in
\cite{BPS} as an attempt to build a healthy extension of HL
gravity in IR (the BPS theory). However the present model differs from the
above theory by an additional condition which fixes the relation between the
scalar field and the aether vector \cite{tedi}. We have also obtained the basic
field equations,
and we have briefly explored the basic cosmological implications of the theory.
A more detailed analysis of the cosmological behavior of the model, as well as
of the astrophysical implications of the theory will be presented in a separate
publication.

\appendix

\section{Proof of the energy-momentum conservation}

In this Section we will derive equation Eq.~\eqref{eq-3} in detail. After taking
the covariant divergence of Eq.~\eqref{eq-1}, and using Eq.~\eqref{eq-2} to
eliminate $dV/d\phi$, we obtain
\begin{align}  \label{d-1}
\kappa^2&\nabla^\mu T_{\mu\nu}=c_1(\nabla^\mu
S_{1\mu\nu}+\nabla_\alpha\nabla_\mu\nabla^\alpha\nabla^\mu\phi\nabla_\nu\phi)
\notag \\
&+c_2(\nabla^\mu S_{2\mu\nu}+\Box^2\phi\nabla_\nu\phi)+c_3\big[\nabla^\mu
S_{3\mu\nu}  \notag \\
&+\nabla_\alpha\Big(\nabla^\mu\phi
\nabla^\rho\phi\nabla_\mu\nabla_\rho\nabla^\alpha\phi+\Box\phi\nabla^\rho%
\phi\nabla _\rho\nabla^\alpha\phi\Big)\nabla_\nu\phi\big]  \notag \\
&+c_4\big[\nabla^\mu S_{4\mu\nu}-\frac{1}{2}\nabla_\alpha(\rho
u^\alpha)\nabla_\nu\phi\big]  \notag \\
&+\nabla^\mu\big[\lambda\nabla_\mu\phi\nabla_\nu\phi-\frac{1}{2}\lambda
g_{\mu\nu}(\nabla_\rho\phi\nabla^\rho\phi+\epsilon)\big]  \notag \\
&-\nabla_\mu(\lambda\nabla^\mu\phi)\nabla_\nu\phi,
\end{align}
where we have defined
\begin{align}  \label{d-2}
S_{1\mu\nu}&=\Box\phi\nabla_\mu\nabla_\nu\phi-2\Box\nabla_{(\mu}
\phi\nabla_{\nu)}
\phi+\nabla^\lambda\phi\nabla_\lambda\nabla_\mu\nabla_\nu\phi
-\frac{1}{2}g_{\mu\nu}
\nabla_\alpha\nabla_\beta\phi\nabla^\alpha\nabla^\beta\phi \\
S_{2\mu\nu}&=g_{\mu\nu}\nabla^\lambda\phi\nabla_\lambda\Box\phi+\frac{1}{2}%
g_{\mu\nu}(\Box\phi)^2-2\nabla_{(\mu}\phi\nabla_{\nu)}\Box\phi \\
S_{3\mu\nu}&=\nabla^\alpha\phi\nabla^\beta\phi\nabla_\mu\nabla_\alpha\phi%
\nabla_\nu\nabla_\beta\phi
-\nabla_\alpha\nabla_\beta\phi\nabla^\alpha\nabla^\beta\phi
\nabla_\mu\phi\nabla_\nu\phi
-\nabla_\alpha\phi\nabla_\beta\nabla^\alpha\nabla^\beta\phi%
\nabla_\mu\phi\nabla_\nu\phi  \notag \\
&\hspace{1cm}-\frac{1}{2}g_{\mu\nu}\nabla^\alpha\phi\nabla^\beta\phi\nabla_%
\alpha\nabla_\lambda\phi\nabla_\beta\nabla^\lambda\phi \\
S_{4\mu\nu}&=\frac{1}{2}T_{\mu\nu}u^{\alpha}\nabla_\alpha\phi.
\end{align}
Using equaions \eqref{eqlam} and \eqref{1}, one can easily prove that the last
two lines of equation \eqref{d-1} is zero.
Expanding the term which corresponds to the constant $c_1$ in Eq.~\eqref{d-1}
results in
\begin{align}  \label{d-4}
2\nabla^\lambda\phi\nabla_{[\mu}\nabla_{\lambda]}\nabla^\mu\nabla_\nu\phi+2%
\nabla^\alpha\nabla^\beta\phi\nabla_{[\beta}\nabla_{\nu]}
+2\nabla_\mu\nabla_\nu\phi\nabla^{[\mu}\nabla^{\alpha]}\nabla_%
\alpha\phi,
\end{align}
which is identically zero due to the relations
\begin{subequations}
\label{d-5}
\begin{align}
&2\nabla_{[\mu}\nabla_{\nu]}T_{\rho\sigma}=T^\alpha_{~\sigma}R_{\alpha\rho%
\nu\mu}+T_\rho^{~\alpha}R_{\alpha\sigma\nu\mu}, \\
&2\nabla_{[\mu}\nabla_{\nu]}A_\rho=A^\alpha R_{\alpha\rho\nu\mu}.
\end{align}
The term corresponding to  constant $c_2$ vanishes by the substitution of the
tensor $S_{2\mu\nu}$. The term corresponding to the constant $c_3$ can be
written as
\end{subequations}
\begin{align}  \label{d-6}
&2\nabla_\nu\phi\nabla_\mu\phi\nabla_\sigma\nabla_\alpha\phi\nabla_{[\alpha}%
\nabla_{\sigma]}\nabla^\sigma\phi
+2\nabla_\nu\phi\nabla^\beta\phi\nabla^\sigma\phi\nabla_{[%
\alpha}\nabla_{\beta]}\nabla_\sigma\nabla^\alpha\phi  \notag \\
&\hspace{1cm}+2\nabla_\nu\phi\nabla_\beta\phi\nabla^\beta\nabla^\alpha\phi%
\nabla_{[\alpha}\nabla_{\sigma]}\nabla^\sigma\phi
+2\nabla^\alpha\phi\nabla^\beta\phi\nabla_\alpha\nabla^\mu\phi%
\nabla_{[\mu}\nabla_{\nu]}\nabla_\beta\phi,
\end{align}
which is zero due to identities \eqref{d-5}. Equation \eqref{d-1} then
reduces to
\begin{align}  \label{d-7}
\kappa^2\nabla^\mu T_{\mu\nu}=c_4\left[\frac{1}{2}\nabla^\mu(T_{\mu\nu}u^{%
\alpha}\nabla_\alpha\phi)-\frac{1}{2}\nabla_\alpha(\rho
u^\alpha)\nabla_\nu\phi\right]
\end{align}
which after expanding and isolating the covariant divergence of the
energy-momentum tensor reduces to equation \eqref{eq-3} in the main text.


\begin{thebibliography}{9}
\bibitem{Szek1} G. Szekeres, Phys. Rev. \textbf{97}, 212 (1955).

\bibitem{Szek2} W. Kantor and G. Szekeres, Phys. Rev. \textbf{104}, 831
(1956).

\bibitem{Szek3} G. Szekeres, Phys. Rev. \textbf{104}, 1791 (1956).

\bibitem{Jac1} T. Jacobson and D. Mattingly, Phys. Rev. \textbf{D 64} 024028
(2001); T. Jacobson and D. Mattingly, Phys. Rev. \textbf{D 70}, 024003
(2004).

\bibitem{Jac2} W. Donnelly and T. Jacobson, Phys. Rev. \textbf{D 82}, 064032
(2010); W. Donnelly and T. Jacobson Phys. Rev. \textbf{D82}, 081501 (2010).

\bibitem{aet} S. M. Carroll and E. A. Lim,
Phys. Rev. {\bf D 70}, 123525 (2004); T. Jacobson, PoSQG-PH 020 (2007); C. Armendariz-Picon, A. Diez-Tejedor, and R. Penco,  JHEP {\bf 1010}, 079 (2010).
\bibitem{hora} P. Horava, Phys. Rev. D 79, 084008 (2009), arXiv:0901.3775
[hep-th].
    \bibitem{inst} C. Charmousis, G. Niz, A. Padilla and P. M. Saffin, JHEP
0908, 070 (2009), arXiv:0905.2579;
    M. Li and Y. Pang, JHEP 0908, 015 (2009), arXiv:0905.2751;
    D. Blas, O. Pujolas and S. Sibiryakov, JHEP 0910, 029 (2009),
arXiv:0906.3046;
    K. Koyama and F. Arroja, JHEP 1003, 061 (2010), arXiv:0910.1998.
    \bibitem{BPS} D. Blas, O. Pujolas and S. Sibiryakov, Phys. Lett. B 688, 350
(2010), arXiv:0912.0550 [hep-th].
    \bibitem{HL} C. Germani, A. Kehagias and K. Sfetsos, JHEP 0909, 060 (2009),
arXiv:0906.1201 [hep-th];
    T. Jacobson, arXiv:1001.4823 [gr-qc].
    \bibitem{black} E. Barausse, T. Jacobson, T. P. Sotiriou,  Phys. Rev. D83
(2011) 124043, arXiv:1104.2889 [gr-qc].
    \bibitem{ted} T. Jacobson, Phys.Rev.D81:101502,2010; arXiv:1001.4823
[hep-th].
    \bibitem{tedi} T. Jacobson, A. J. Speranza, arXiv:1405.6351 [gr-qc].
\bibitem{MOND} M. Milgrom,  ApJ {\bf 270}, 365 (1983);
M. Milgrom, ApJ {\bf 332}, 86  (1988);
M. Milgrom,  Phys. Rev. Lett. {\bf 109}, 131101 (2012).

\bibitem{Bek} J. D. Bekenstein, Phys. Rev. {\bf D70}, 083509 (2004).

\bibitem{Zlos} T. G. Zlosnik, P. G. Ferreira, and G. D. Starkman, Phys. Rev. {\bf D 74},  044037 (2006); C. Skordis and  T. G. Zlosnik, Phys. Rev. {\bf D 85}, 044044 (2012).

    \bibitem{aet1} C. Eling and T. Jacobson, Class. Quant. Grav. {\bf 23}, 5625 (2006); Erratum-ibid. {\bf 27}, 049801 (2010).

    \bibitem{aet2} C. Eling and T. Jacobson, Class. Quant. Grav. {\bf 23}, 5643 (2006); Erratum-ibid. {\bf 27}, 049802 (2010).
    \bibitem{aet3} R. A. Konoplya and A. Zhidenko, Phys. Lett. {\bf B 648}, 236 (2007).
    \bibitem{aet4} D. Garfinkle, C. Eling, and T. Jacobson, Phys. Rev. {\bf D 76}, 024003 (2007); C. Eling, T. Jacobson, and M. C. Miller, Phys. Rev. {\bf D 76}, 042003 (2007); Erratum-ibid. {\bf D 80}, 129906 (2009); B. Z. Foster, Phys. Rev. {\bf D 76}, 084033 (2007); T. Tamaki and U. Miyamoto, Phys. Rev. {\bf D 77}, 024026 (2008); C. Gao and Y.-G. Shen, Phys. Rev. {\bf D 88}, 103508 (2013).
        \bibitem{aet5} C. Bonvin, R. Durrer, P. G. Ferreira, G. Starkman, and T. G. Zlosnik, Phys. Rev. {\bf D 77}, 024037 (2008); Y. Xie and T.-Y. Huang, Phys. Rev. {\bf D 77}, 124049 (2008); D.-C. Dai, R. Matsuo, and G. Starkman, Phys. Rev. {\bf D 78}, 104004 (2008); D. Garfinkle, J. Isenberg, and J. M. Martin-Garcia, Phys. Rev. {\bf D 86},  084009 (2012).
            \bibitem{aet5a} P. Berglund, J. Bhattacharyya, and D. Mattingly,
Phys. Rev. Lett. {\bf 110}, 071301 (2013).
                \bibitem{aet6a} J. A. Zuntz, P. G. Ferreira, and T. G. Zlosnik,
Phys. Rev. Lett. {\bf 101}, 261102 (2008).
\bibitem{24} A. H. Chamseddine, V. Mukhanov and A. Vikman, JCAP 1406 (2014)
017; E. A. Lim, I. Sawicki and A. Vikman, JCAP 1005 (2010) 012.
\bibitem{aet7} C. Armendariz-Picon, N. F. Sierra, and J. Garriga, Journal of Cosmology and Astroparticle Physics {\bf 07}, 010 (2010).
\bibitem{aet8} M. Nakashima and T. Kobayashi, Phys. Rev. {\bf D 84}, 084051 (2011).
\bibitem{aet9} J. D. Barrow, Phys. Rev. {\bf D 85}, 047503 (2012).
\bibitem{aet10} X. Meng and X. Du, Phys. Lett. {\bf B 710}, 493 (2012).
\bibitem{aet11} P. Sandin, B. Alhulaimi, and A. Coley, Phys. Rev. {\bf D 87}, 044031 (2013).
\bibitem{aet12} B. Alhulaimi, A. Coley, and P. Sandin, Journal of Mathematical Physics {\bf 54},  042503-042503-29 (2013).
\bibitem{aet13} H. Wei, X.-P. Yan, and Y.-N. Zhou, Gen. Rel. Grav. 46, 1719 (2014).
\bibitem{aet14} A. R. Solomon and J. D. Barrow, Phys. Rev. {\bf D 89}, 024001 (2014).

\bibitem{22fRT} T. Harko, Phys. Lett. {\bf B 669}, 376 (2008); T. Harko, F. S.N. Lobo, S. Nojiri, and S. D. Odintsov, Phys. Rev. {\bf D 84}, 024020 (2011).
\bibitem{Bert} O. Bertolami, C. G. Bohmer, T. Harko, and F. S.N. Lobo, Phys. Rev. {\bf D 75}, 104016 (2007).

\end{thebibliography}
\end{document}